\documentclass{PoS}

\title{Tau decays and dark-sector searches at BaBar}

\ShortTitle{Tau decays and dark-sector searches at BaBar}

\author{\speaker{Benjamin Oberhof}\thanks{On behalf of the BaBar collaboration.}\\
        Laboratori Nazionali di Frascati dell'INFN\\
        E-mail: \email{benjamin.oberhof@lnf.infn.it}}

\abstract{We present some recent measurements performed using 514 fb$^{-1}$ of $e^+e^-$ collisions collected with the BaBar detector at the PEP-II asymmetric collider at SLAC.  
First we present a study of the decay $\tau^{\pm} \rightarrow K^{\pm} K_S \nu_{\tau}$: we measure the branching fraction B($\tau^{\pm} \rightarrow K^{\pm} K_S \nu_{\tau}$), 
the mass spectrum of the $K^{\pm}K_S$ system and the spectral function. Our results are in agreement with and improve upon previous measurements. 
We then report a search for a dark boson $A'$ produced in the reaction $e^+ e^- \rightarrow \gamma A'$ in invisible final states 
$A' \rightarrow \chi \bar \chi$.  We find no evidence for such processes and set 90\% confidence level upper limits on the coupling strength 
of $A'$ as a function of the $A'$ mass $m_{A'}$. In particular, our limits exclude the values of the $A'$ coupling suggested by the dark-photon interpretation of the muon $(g-2)$ anomaly. 
Finally we report the search for a light non-Standard Model gauge boson $Z'$ coupling only to the second and third lepton families. 
Our results significantly improve current limits and further constrain the remaining region of the allowed parameter space.}

\FullConference{XIV International Conference on Heavy Quarks and Leptons (HQL2018)\\
		May 27- June 1, 2018\\
		Yamagata Terrsa, Yamagata, Japan}

\begin{document}

\section{Introduction}

We present recent results obtained using data recorded by the BaBar detector at the PEP-II asymmetric-energy $e^+e^-$ storage rings operated at the SLAC National Accelerator Laboratory. 
The data sample consists of 424 fb$^{-1}$ of $e^+e^-$ collisions recorded at the center-of-mass (CM) energy of the $\Upsilon(4S)$ resonance, 
27.8 fb$^{-1}$ of data recorded at the $\Upsilon(3S)$, 13.6 fb$^{-1}$ of data collected at the $\Upsilon(2S)$ resonance and 48 fb$^{-1}$ off-resonance data \cite{lumi_paper}. 
%The $e^+e^-$ CM frame was boosted relative to the detector approximately along the detector?s magnetic field axis by $\beta \gamma \simeq 0.5$. 
A detailed description of the BaBar detector is given elsewhere \cite{detector}. 

\section{Measurement of the spectral function in $\tau^{-} \rightarrow K^{-} K_S \nu_{\tau}$ decays}

With a production cross section of $0.92$ nb at the CM energy of the $\Upsilon (4S)$ \cite{physics}, roughly $10^9$ $\tau$ leptons have been produced at BaBar. 
This large data sample offers the opportunity for high precision measurements in $\tau$ decays. 
Herein we present a measurement of $B(\tau^{-} \rightarrow K^{-} K_S \nu_{\tau})$ and of the spectral function of this decay defined as \cite{spec_fun}
\begin{equation}
V(q) = \frac{m^ 8 _{\tau}}{12 \pi C(q) |V_{ud}| ^2} \frac{B(\tau^- \rightarrow K^- K_S \nu_{\tau})}{B(\tau^- \rightarrow e^-  \nu_{\tau} \bar \nu_e)} \frac{1}{N} \frac{dN} {dq}, 
\label{s_f}
\end{equation}
where $m_{\tau}$ is the $\tau$ mass, $q = m_{K^- K_S}$ is the invariant mass of the $K^- K_S$ system, 
$V_{ud}$ is an element of the CKM matrix, $(dN/dq)/N$ is the normalized $K^- K_S$ mass spectrum, and $C(q)$ is the phase space correction factor given by
\begin{equation}
C(q) = q(m^2 _{\tau} - q^2)^2 (m^2 _{\tau} + 2 q^2).
\end{equation}
Under the conserved-vector-current hypothesis, Eq. \ref{s_f} is related to the isovector part of the $e^+ e^- \rightarrow K^+ K^-$ cross section 
\begin{equation}
\sigma^{I=1} _{e^+ e^- \rightarrow K^+ K^-} (q) = \frac { 4 \pi^2 \alpha^2} {q^2} V (q),
\end{equation} 
where $\alpha$ is the fine structure constant. 
We select $e^+e^- \rightarrow \tau^+ \tau^-$ signal events in which $\tau^- \rightarrow K^- K_S \nu_{\tau}$ and the $\tau^+$ decays leptonically ($\tau^+ \rightarrow l^+ \nu_l \bar \nu_{\tau}$, $l = e$ or $\mu$) 
(and charge conjugated decays). The $K_S$ candidate is reconstructed in the $K_S \rightarrow \pi^+ \pi^-$ decay mode. 
To select the events we require four charged tracks and zero net charge, one identified lepton (electron or muon) and an identified kaon of opposite charge. 
Lepton candidates must satisfy $p_l > 1.2$ GeV/$c$ in the laboratory frame, and $p_{l,CM}<4.5$ GeV/$c$ in the CM. %and $| \cos \theta_l | < 0.9$. 
The $K^-$ candidate must have $0.4 < p_K < 5$ GeV/$c$, and the cosine of its polar angle must lie between $-0.7374 < \cos \theta_K < 0.9005$. 
The invariant mass of the two remaining pion candidates must lie within 25 MeV/c$^2$ of the nominal $K_S$ mass and the $K_S$ flight length $r_{K_S}$, must be larger than 1 cm. 
The total energy in neutral clusters, $E_{tot}$, must be less than 2 GeV and magnitude of the thrust \cite{thrust} for the event, must be greater than 0.875.
Finally the angle defined by the momentum of the lepton and that of the $K^- K_S$ system in the CM must be larger than 110 degrees. 
The selection criteria suppress the $\tau$ background by 3.5 orders of magnitude and the $q \bar q$ background by 5.5. 
Background events surviving the selection consists of events with the decay $\tau^- \rightarrow K^- K_S \pi^0 \nu _{\tau}$ (79\%), 
events with a misidentified kaon from $\tau^- \rightarrow \pi^- K_S \nu_{\tau}$ (10\%) and $\tau^- \rightarrow \pi^- K_S \pi^0 \nu_{\tau}$ (3\%), 
and events with a misidentified lepton from $\tau^+ \rightarrow \pi^+ \bar \nu_{\tau}$ and $\tau^+ \rightarrow \pi^+ \pi^0 \bar \nu _{\tau}$ (7\%). 
The detection efficiency obtained after applying the selection criteria is calculated using signal MC simulation as a function of 
the true $m_{K^- K_S}$ mass and is weakly dependent on $m_{K^- K_S}$. The average over the $m_{K^- K_S}$ mass range is $\simeq$ 13\%. 

\begin {figure}[ht]
\begin{center}
\centering
%  \begin{minipage}[b]{0.49\textwidth}
%  \includegraphics[width=\textwidth]{invisible1.pdf}
 %\end{minipage}
 % \hfill
\begin{minipage}[b]{0.98\textwidth}
\includegraphics[width=1.0\textwidth]{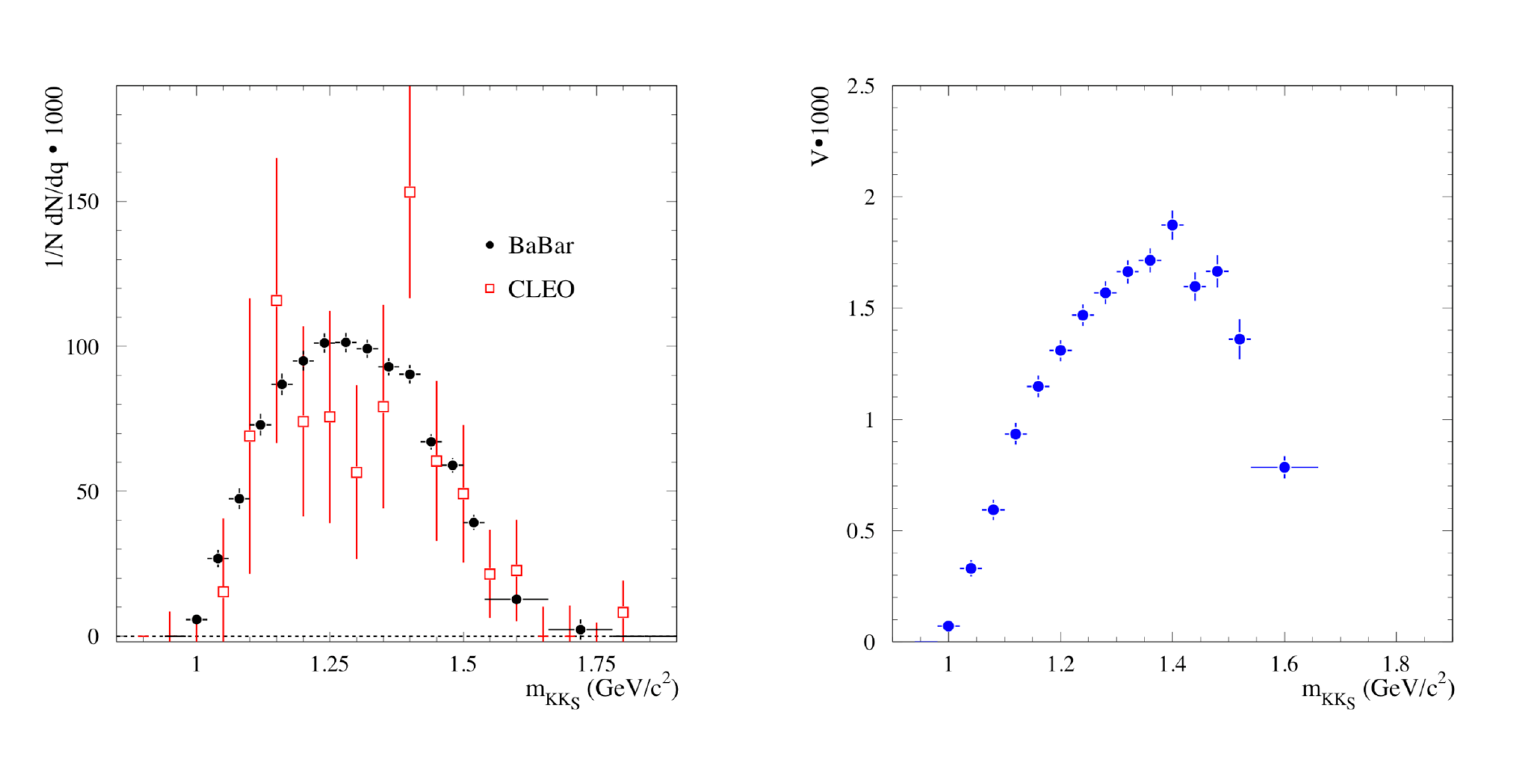}
\end{minipage}
\caption{$K^- K_S$ invariant mass spectrum (left) measured by BaBar (circles) compared to the CLEO's measurement \cite{cleo} (squares) (only statistical uncertainties are shown) and 
the spectral function measured by BaBar (right). Because of the large error in the mass interval 1.66-1.78 GeV/$c^2$, which exceeds the scale $y$-axis, the value of $V(q)$ in this interval is not shown.}
\label{tau_res}
\end{center}
\end{figure}

The branching ratio of the $\tau^- \rightarrow K^- K_S \nu_{\tau}$ decay is obtained as
\begin{equation}
B(\tau^- \rightarrow K^- K_S \nu_{\tau})  = \frac{N_{exp}} { 2 L B_{lep} \sigma_{\tau \tau}} = (0.739 \pm 0.011 \pm 0.020) \times 10^{-3},
\end{equation}
where $N_{exp}$ is the total number of signal events in the spectrum, $L$ is the integrated luminosity, $\sigma_{\tau \tau}$ is the $e^+e^- \rightarrow \tau^+ \tau^-$ cross section and $B_{lep}$ 
is the sum of electronic and muonic branching fractions \cite{pdg}. This result is in good agreement and has comparable precision to existing measurement of $B(\tau^- \rightarrow K^- K_S \nu_{\tau})$ \cite{belle}. 
The measured mass spectrum $m_{K^- K_S}$ is shown in Fig. \ref{tau_res} (left) and compared to the CLEO measurement \cite{cleo}. The two results are in good agreement; however the BaBar 
measurement is far more precise. The spectral function $V(q)$ calculated using Eq. \ref{s_f} is shown in Fig. \ref{tau_res} (right). This result represents the first measurement of $V(q)$. 

\section{Search for the dark photon in invisible final states}

\begin {figure}[ht]
\begin{center}
\centering
%  \begin{minipage}[b]{0.49\textwidth}
%  \includegraphics[width=\textwidth]{invisible1.pdf}
 %\end{minipage}
 % \hfill
\begin{minipage}[b]{0.55\textwidth}
\includegraphics[width=1.0\textwidth]{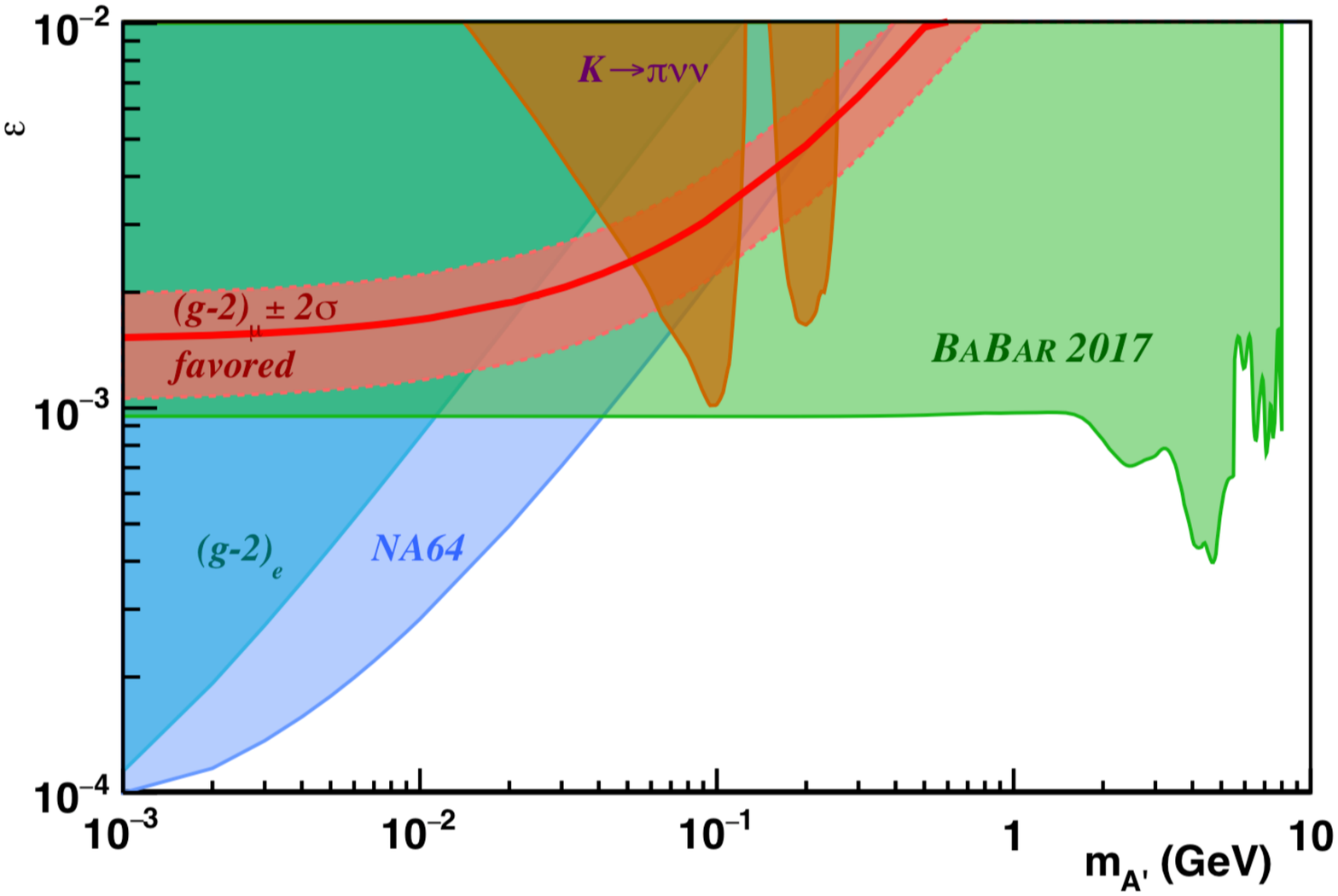}
\end{minipage}
\caption{Upper limits at 90\% CL on the mixing strength $\epsilon$ as a function of $m_{A'}$ excluded by this work compared to the previous constraints as well as the region preferred by the $(g - 2)_{\mu}$ anomaly. }
\label{invisible_res}
\end{center}
\end{figure}

We search for the process $e^+ e^- \rightarrow \gamma A'$, followed by invisible decays of the $A' \rightarrow \chi \bar \chi $ in 53 fb$^{-1}$ of data collected in 2007-2008 with 
CM energies near the $\Upsilon(2S)$, $\Upsilon(3S)$, and $\Upsilon(4S)$ resonances with a special ``single photon'' trigger. 
The $A'$ is supposed to be produced via kinetic mixing with the standard model photon with coupling $\epsilon$ and mass $A'$. 
The signal signature would thus be a peak in the missing mass $M_X ^2 = s - 2 E_{\gamma} \sqrt s/c^2$, where $s$ is the square of the CM energy. % and the aster- isk hereafter denotes a CM quantity.
%Since the production of the A? is not expected to be enhanced by the presence of the ? resonances, we combine the datasets collected in the vicinity of each ? resonance. 
%In order to properly account for different efficiency as a function of the CM energy, we measure the signal event yields separately for the three different datasets. 
Two different single-photon trigger lines were used for this search the high-energy photon line (``LowM''), which accepts events with a cluster in the electro-magnetic calorimeter (EMC) with energy $E > 2$ GeV in the CM frame 
and no tracks originating from the interaction region (IR), and a low-energy line (``HighM''), which requires an EMC cluster with energy $E > 1$ GeV and no tracks originating from the IR. 
The total data sample collected with the ``LowM'' triggers is 53 fb$^{-1}$ while the total data sample collected with the ``HighM'' triggers is 35.9 fb$^{-1}$. 
We then tighten the selection requiring $E_{\gamma} > 3$ GeV and no drift chambers (DCH) tracks with CM momentum $p > 1$ GeV/$c$ for the ``LowM'' selection and 
$E_{\gamma} > 1.5$ GeV/$c$ and no DCH tracks with momentum $p > 0.1$ GeV/$c$ for the ``HighM'' selection. 
This selection naturally splits the dataset into two broad $M_X$ ranges: the ``LowM'' selection is used for the low $M_X$ region $-4 < M_X ^2 < 36$ GeV$^2/c^4$ 
where the dominant background is the QED process $e^+e^- \rightarrow \gamma \gamma$, peaking at $M_X \simeq 0$ ($E_{\gamma}  \simeq \sqrt s/2$). 
Because of the orientation of the EMC crystals, which point towards the IR, one of the photons may escape detection even if it is within the nominal EMC acceptance. 
%The event selection is optimized to reduce this peaking background as much as possible. 
The ``HighM'' trigger selection defines the high $M_X$ range $24 < M_X ^2 < 69$ GeV$^2/c^4$; this region is dominated by the low-angle radiative Bhabha 
events $e^+e^- \rightarrow e^+e^- \gamma$, in which both the electron and the positron are outside the detector acceptance. 
We require photon candidates to have $| \cos \theta | < 0.6$ to reject radiative Bhabha events that peak in 
the forward and backward directions. We also require the event to contain no charged particle tracks. 
Further selection is obtained using a multivariate boosted decision tree (BDT) \cite{bdt} discriminant based on 12 different discriminating variables. 
The BDTs are trained separately in ``LowM'' and ``HighM'' regions. 
Each BDT is trained using an equal number of simulated signal events, with uniformly distributed $A'$ masses, 
and background events from the $\Upsilon(3S)$ on-peak sample. 
The final selection is defined using a statistically independent sample with the same composition as those used for BDT training and is optimized 
to minimize the expected upper limit on the cross section $\sigma _{A'}$ of $e^+e^- \rightarrow \gamma A'$. 
These samples are used also to measure signal efficiency. 
The cross section $\sigma _{A'}$ is measured as a function of the assumed mass $m_{A'}$  by performing a series of un-binned extended maximum likelihood fits 
to $M_X ^2$; $m_{A'}$ is varied in steps roughly equal to half of the mass resolution. A set of simultaneous fits to $\Upsilon(2S)$, $\Upsilon(3S)$, and for the ``LowM'' region $\Upsilon(4S)$ datasets is performed. 
The signal PDF is described by a Crystal Ball function \cite{crystalball} around the nominal value $m^2 _{A'}$ while the background PDF is given by the sum of two components 
a peaking contribution from $e^+e^- \rightarrow \gamma \gamma$ events described by a Crystal Ball function, 
and a non-peaking contribution from $e^+e^- \rightarrow \gamma e^+e^-$ which is described by a second order polynomial for $m_{A'} \le 5.5$ GeV$/c^2$ 
and a sum of exponentiated polynomials for $5.5 < m_{A'} \le 8.0$ GeV$/c^2$. 
The largest contribution to the systematic error in the signal yield are from the shape 
of the signal and background PDFs, and the uncertainties in the efficiency of signal and trigger selections. 
The most significant deviation of $\epsilon$ from zero occurs for $m_{A'} = 6.21$ GeV$/c^2$ and corresponds to a local significance $S$ = 3.1, where $S = 2 \ln(L_{max} /L_0 )$, 
$L_{max}$ is the maximum value of the likelihood, and $L_0$ is the value of the likelihood with the signal yield fixed to zero. 
The probability to find such a deviation in any of the 166 $m_{A'}$ points in the absence of signal is $\simeq$ 1\%, corresponding to a significance of $2.6 \sigma$. 
The 90\% confidence level (CL) upper limit on $\epsilon$ as a function of $m_{A'}$ is shown in Fig. \ref{invisible_res}. 
Our results \cite{dark_paper} exclude the dark-photon coupling as the explanation for the $(g - 2)_{\mu}$ \cite{gminus2} anomaly and place stringent constraints 
over a broad range of parameter space, significantly improving previous results. 
This search complements previous searches carried out at BaBar for visible dark photon decays \cite{abi}-\cite{visible}. 

\section{Search for a muonic dark force}

SM particles and interactions are insufficient to explain cosmological and astrophysical observations of dark matter. 
A possible scenario is represented by new hidden sectors that are only feebly coupled to the SM. 
In the simplest case of a hidden U(1) gauge interaction, such a sector contains its own gauge bosons $Z'$ and 
SM fields may directly couple to the $Z'$, or alternatively the $Z'$ boson may mix with the SM hypercharge boson \cite{holdom}. 
In the latter case, the $Z'$ couplings are proportional to the SM gauge couplings; however, due to large couplings to electrons and light-flavor quarks, 
such scenarios are strongly constrained by existing searches \cite{anastasi2}. 
If SM fields are directly charged under the dark force instead, the $Z'$ may interact preferentially with heavy-flavor leptons, 
greatly reducing the sensitivity of current searches. Such interactions could account for the experimentally measured value of the 
muon anomalous magnetic dipole moment \cite{pospelov}, as well as the discrepancy in the proton radius extracted from measurements of the 
Lamb shift in muonic hydrogen compared to observations in non-muonic atoms \cite{barger, tucker}. 
In the following we report a search for dark bosons $Z'$ with vector couplings only to the second and third generations of leptons \cite{he} 
in $e^+e^- \rightarrow \mu^+ \mu^- Z'$, $Z' \rightarrow \mu^+ \mu^-$. 
For this search we use the full BaBar data sample. % plus about 28 fb$^{-1}$ data at $\Upsilon (3S)$, 14 fb$^{-1}$ data at $\Upsilon (2S)$, and 48 fb$^{-1}$ off-resonance data. 
About 5\% of the data set is used to validate and optimize the analysis method; the rest of the data was only examined after finalizing the analysis method. 
For the background study we use MC samples. Signal MC events are generated using MadGraph 5 \cite{madgraph} and hadronized in Pythia 6 \cite{pythia} 
for $Z'$ mass hypotheses from the dimuon mass threshold to 10.3 GeV/$c^2$. 
%Background samples include the direct processes of $e^+e^- \rightarrow \mu^+\mu^-\mu^+\mu^-$ generated with Diag36 \cite{diag36}, 
%which includes the full set of the lowest order diagrams. 
%The events of the process of $e^+e^- \rightarrow e^+e^- (\gamma)$ are generated using BHWIDE \cite{bhwide} while of $e^+e^- \rightarrow \mu^+ \mu^- (\gamma)$  
%and $e^+e^- \rightarrow \tau^+ \tau^- (\gamma)$ are generated using KKMC \cite{kkmc}. 
%The off-resonance data samples, $e^+e^- \rightarrow q \bar q$ ($q$ = $u$, $d$, $s$, $c$), are simulated using JETSET. 
%The events processes of $e^+e^- \rightarrow \psi(2S) \gamma$ followed by $\psi (2S) \rightarrow \pi^+ \pi^- J/\psi$ and $J/\psi \rightarrow \mu^+ \mu^-$ 
%were generated using EvtGen \cite{evtgen} with appropriate phase-space model. 
%Finally the detector acceptance and reconstruction efficiency are determined using GEANT4.
%\begin{figure}[htb]
%\centering
%\includegraphics[width=0.49\textwidth, clip]{muonic_inv_m.pdf}
%\caption{The distribution of the four-muon invariant mass, $m_{4\mu}$, for data taken at the $\Upsilon (4S)$ peak together with Monte Carlo predictions of various processes 
%normalized to data luminosity. The $e^+ e^- \rightarrow \mu^+ \mu^- \mu^+ \mu^-$ MC does not include ISR corrections.}
%\label{muonic_inv_m}
%\end{figure}
We select events with exactly two pairs of oppositely charged tracks, consistent with the $e^+e^- \rightarrow \mu^+ \mu^- Z'$, $Z' \rightarrow \mu^+ \mu^-$ final state. 
The muons are identified by multivariate particle identification algorithms for each track. 
We require the sum of energies of the deposits in the electromagnetic calorimeter that are not associated to any track to be less than 200 MeV. 
We reject events coming from the $\Upsilon (3S)$ and $\Upsilon (2S)$, where $\Upsilon(2S, 3S) \rightarrow \pi^+ \pi^- \Upsilon (1S)$, $\Upsilon (1S) \rightarrow \mu^+ \mu^-$ 
if the dimuon combination lies within 100 MeV/$c^2$ of the $\Upsilon(1S )$ mass. 
%The distribution of the four-muon invariant mass after all selections is shown in Fig.~\ref{muonic_inv_m}. 
The lower part of the four-muon invariant mass spectrum, $m_{4\mu} < 9$ GeV/$c^2$, is well reproduced by the Monte Carlo simulation 
while the MC simulation overestimates the full energy peak by about 30\% and fails to reproduce the radiative tail. 
This, however, is expected because the Diag36 \cite{diag36} simulation used to simulate $e^+e^- \rightarrow \mu^+\mu^-\mu^+\mu^-$ does not simulate the initial state radiation (ISR). 
We select $e^+e^- \rightarrow \mu^+ \mu^- \mu^+ \mu^-$ events by requiring a four-muon invariant 
mass distribution within 500 MeV/$c^2$ of the nominal center-of-mass energy. We also require the tracks to originate 
from the interaction point to within its uncertainty and constraining the center-of-mass energy of the system to be within the beam energy spread. 
We do not attempt to select a single $Z' \rightarrow \mu^+ \mu^-$ candidate per event, but instead consider all possible combinations.
The most important contribution on the invariant mass peak besides the QED process $e^+ e^- \rightarrow \mu^+ \mu^- \mu^+ \mu^-$ comes from 
$\Upsilon(2S) \rightarrow \pi^+ \pi^- J/\psi$, $J/\psi \rightarrow \mu^+ \mu^-$. %as can be seen in Fig.~\ref{muonic_inv_m}. 
We extract the signal yield by a series of unbinned likelihood fits to the spectrum of the reduced dimuon mass $m_R$, 
defined as $m_R = \sqrt{m^2_{\mu^+ \mu^-} - 4 m^2_{\mu}}$, within the range of $0.212 < m_R < 10$ GeV$/c^2$ and $0.212 < m_R < 9$ GeV$/c^2$ 
for the $\Upsilon (4S)$ resonance data and $\Upsilon(2S)$ and $\Upsilon(3S)$ resonances data, respectively. 
We exclude a region of $\pm30$ MeV/$c^2$ around the nominal $J/\psi$ mass. We probe a total of 2219 mass hypotheses. 
The signal efficiency at low masses is about 35\% and rises to about 50\% around $m_R=6-7$ GeV/$c^2$ to dropping again at higher values of the reduced dimuon mass. 
%The signal efficiencies include a correction factor of 0.82, which accounts for the impact of ISR not included in the simulation, 
%as well as differences between data and simulation in trigger efficiency, charged particle identification, and track and photon reconstruction efficiencies. 
%This correction factor is obtained from the ratio of the $m_R$ distribution in simulated $e^+ e^- \rightarrow \mu^+ \mu^- \mu^+ \mu^-$ events to the observed distribution 
%in the mass region 1-9 GeV$/c^2$. 
%We also assign a systematic uncertainty of 5\% to cover the small variations between the uncertainties on the $e^+e^- \rightarrow \mu^+ \mu^- \mu^+ \mu^-$ and data taking period. 
%We calculate the ISR contribution based on the quasi real electron approximation \cite{batell}. 
\begin {figure}[ht]
\begin{center}
  \centering
  \begin{minipage}[b]{0.49\textwidth}
  \includegraphics[width=\textwidth]{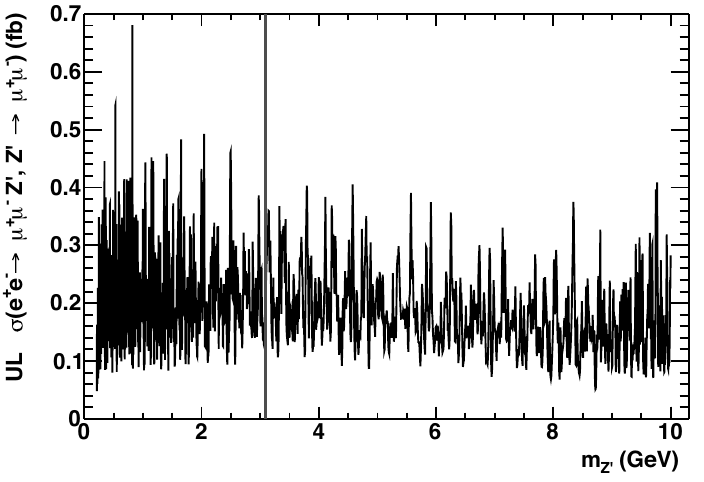}
 \end{minipage}
  \hfill
  \begin{minipage}[b]{0.49\textwidth}
\includegraphics[width=\textwidth]{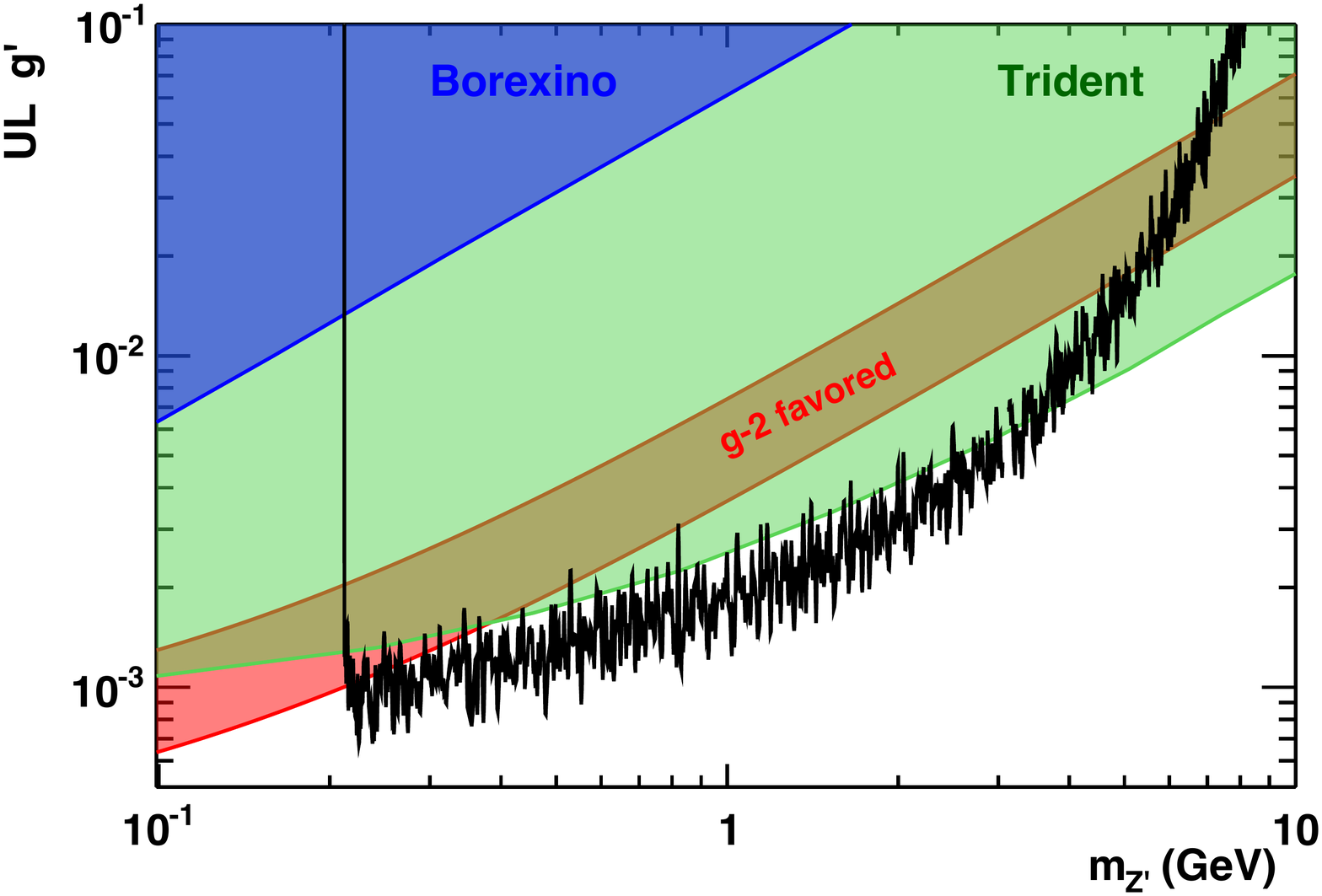}
  \end{minipage}
\caption{The 90\% CL upper limits on the cross-section $\sigma (e^+e^- \rightarrow \mu^+ \mu^- Z'$, $Z' \rightarrow \mu^+ \mu^-$) as a function of the $Z'$ mass (left) and 
the 90\% CL upper limits on the new gauge coupling $g'$ as a function of the $Z'$ mass, together with the constraints derived from the production of a $\mu^+ \mu^-$ pairs in $\nu_{\mu}$ 
scattering (``Trident'') \cite{trident} (right). The region consistent with the discrepancy between the calculated and measured anomalous magnetic moment of the muon 
within 2$\sigma$ is shaded in red.}
     \label{muonic_res}
\end{center}
\end{figure}
The cross section of $e^+e^- \rightarrow \mu^+ \mu^- Z'$, $Z' \rightarrow \mu^+ \mu^-$ is extracted as a function of $Z'$ mass. 
The black band at $\sim$ 3.1 GeV/$c^2$  indicates the excluded region. We find the largest local significance is 4.3$\sigma$ around a $Z'$ mass of 0.82 GeV/$c^2$ 
that corresponds to a global significance of $1.6\sigma$; this is consistent with the null-hypothesis \cite{bertrand}. 
We also derive 90\% confidence level (CL) Bayesian upper limit on the cross section of $e^+e^- \rightarrow \mu^+ \mu^- Z'$, $Z' \rightarrow \mu^+ \mu^-$  as shown in Fig.~\ref{muonic_res} (left).
We consider all uncertainties to be uncorrelated except for the uncertainties of the luminosity and efficiency. 
Finally, we extract the corresponding 90\% CL on the coupling parameter $g'$ by assuming the equal 
magnitude vector couplings muons, taus and the corresponding neutrinos together with the existing limits from 
Borexino \cite{kamada} and neutrino experiments as shown in Fig.~\ref{muonic_res} (right).

\end{document}